\documentclass[aps,prl,twocolumn,showpacs,preprintnumbers,floatfix]{revtex4}
\usepackage{eurosym}
\usepackage{amsmath}
\usepackage{amssymb}
\usepackage{epsfig}

\begin{document}

\title{Excitation and control of large amplitude standing magnetization waves%
}
\author{L. Friedland}
\email{lazar@mail.huji.ac.il}
\affiliation{Racah Institute of Physics, Hebrew University of Jerusalem, Jerusalem 91904,
Israel}
\author{A. G. Shagalov}
\email{shagalov@imp.uran.ru}
\affiliation{Institute of Metal Physics, Ekaterinburg 620990, Russian Federation}
\affiliation{Ural Federal University, Mira 19, Ekaterinburg 620002, Russian Federation}

\begin{abstract}
A robust approach to excitation and control of large amplitude standing
magnetization waves in an easy axis ferromagnetic by starting from a ground
state and passage through resonances with chirped frequency microwave or
spin torque drives is proposed. The formation of these waves involves two
stages, where in the first stage, a spatially uniform, precessing
magnetization is created via passage through a resonance followed by a
self-phase-locking (autoresonance) with a constant amplitude drive. In the
second stage, the passage trough an additional resonance with a spatial
modulation of the driving amplitude yields transformation of the uniform
solution into a doubly phase-locked standing wave, whose amplitude is
controlled by the variation of the driving frequency. The stability of this
excitation process is analyzed both numerically and via Whitham's averaged
variational principle.
\end{abstract}

\pacs{75.-n,75.78.Fg}
\maketitle

\section{I. Introduction}

Because of the complexity and despite decades of studies, magnetization
dynamics in ferromagnetic materials remains of interest to basic and applied
research. For example, nonlinear spin waves and solitons in ferromagnetic
films were studied experimentally extensively (e.g. \cite{Scott1,Scott2,
Patton1, Patton2}). Magnetostatic and boundary effects in such macroscopic
films yield complex dispersion of the spin waves. Depending on the sign of
the dispersion both bright and dark magnetic solitons were observed. The
long wavelength approximation in this problem yields the nonlinear
Schrodinger (NLS) model, providing a convenient theoretical basis for
investigation. The NLS equation has well known traveling wave and soliton
solutions, \cite{Scott}, allowing interpretation of the experimentally
observed magnetization dynamics.

In recent years, applications in ferromagnetic nanowires opened new
perspectives in studying magnetization waveforms \cite{Braun}. At the
nanoscales a quasi-one-dimensional symmetry can be realized and
magnetostatic effects can be reduced to additional contributions to the
anisotropy \cite{Braun,Garbou,Kohn}, which can be conveniently modeled by
the Landau-Lifshitz-Gibert (LLG)\ equation. It is known that the
one-dimensional (1D), dissipationless LLG equation, similar to the NLS
equation is integrable and has a multitude of exact solutions including
solitons and spatially periodic waveforms \cite{Braun,Kosevich,M}, expected
to be observed in nanowires. The simplest solitons are domain walls, which
are studied extensively \cite{Beach,R0,Neg,Sitte,W} as a basis for new
memory and logic devices \cite{Cr,Th}. A different type of solitons are
so-called breathers \cite{Braun}, which can be interpreted as an interacting
pair of domain walls with opposite topological charges (soliton-antisoliton
pair). They are stable localized objects in easy-axis ferromagnetic when
dissipation is negligible \cite{Kosevich}, which was also illustrated in
numerical simulations \cite{Kosevich2}. These breather solitons correspond
to the bright NLS solitons in the small amplitude approximation \cite%
{Kosevich}. Solitons in ferromagnetic nonowires with a spin polarized
current were also discussed in \cite{Li1,Li2} in the framework of a modified
NLS model.

In this work, we focus on excitation of large amplitude standing LLG waves
in an easy axis ferromagnetic, such that the projection $M_{z}$ of the
magnetization vector $\mathbf{M}$ on the easy axis is independent of time
and periodic in $z$, while $\mathbf{M}_{\bot }$ precesses uniformly around
the axis. These waves approach a soliton limit as their wavelength increases
(see below). The question is how to generate such waves by starting from a
\textit{simple} initial equilibrium and how to control their dynamics?
Excitation by an impulse or localized external fields usually are unsuitable
for generating pure large amplitude standing waves because of significant
residual perturbations. Here, we suggest a simple method of exciting these
waves based on the autoresonance approach via driving the system by a small,
chirped frequency external rotating magnetic field or a spin torque. This
approach allows to excite the waves with a predefined amplitude and phase
and stabilize them with respect to dissipation. The autoresonance approach
uses the salient property of a nonlinear system to stay in resonance with
driving perturbations despite slow variation of parameters. The idea was
used in many applications starting from particle accelerators \cite%
{McMillan,Veksler}, through planetary dynamics \cite{Malhotra}, \cite%
{Planetary} and atomic physics \cite{Atomic1,Atomic2}, to plasmas \cite%
{Plasma}, magnetization dynamics in single domain nanoparticles \cite%
{Manfredi,Lazar145,Lazar152}, and more. Autoresonant excitation of both
bright and dark solitons and spatially periodic multiphase waves within the
NLS model were studied in Ref. \cite{Lazar82,L2,Lazar104}, while the
autoresonant control of NLS solitons is described in Refs. \cite{Batalov,S2}%
. In all these applications, one drives the system of interest by an
oscillating perturbation, captures it into a nonlinear resonance, while
slowly varying the driving frequency (or other parameter). The resulting
continuing self-phase-locking (autoresonance) yields excursion of the system
in its solutions space, frequently leading to emergence and control of
nontrivial solutions. In this work, motivated by the aforementioned results
in related driven-chirped NLS systems, we apply a similar approach yielding
arbitrary amplitude, standing magnetization waves.

The scope of the presentation will be as follows. In Sec. II, we introduce
our autoresonant magnetization model and discuss the problem of capturing
the system into resonance with a chirped frequency microwave field followed
by formation of an autoresonant, spatially uniform magnetization state. In
Sec. III, we study transition from the uniform state to a standing wave by
spatially modulating the amplitude of the chirped frequency drive. In the
same section, we will illustrate this process in simulations and present a
qualitative picture of the dynamics. Section IV will be focussed on the
theory of the autoresonant standing waves and discuss their modulational
stability via Whitham's averaged variational principle \cite{Whitham}. In
Sec. V, we illustrate excitation of the standing waves via spin torque
driving and, finally, Sec. VI will present our conclusions.

\section{II Autoresonant magnetization model}

Our starting point is the\ $1D$ Landau-Lifshitz-Gilbert (LLG) equation for a
ferromagnetic with the easy axis along $\widehat{\mathbf{e}}_{z}$ in an
external magnetic field $\mathbf{H}=H_{0}\widehat{\mathbf{e}}_{z}$\ and in
the presence of a weak rotating driving microwave field $\mathbf{H}%
_{d}=a(\cos \varphi _{d}\widehat{\mathbf{e}}_{x}+\sin \varphi _{d}\widehat{%
\mathbf{e}}_{y})$ of constant amplitude and slowly chirped frequency $\omega
_{d}(t)=-\partial \varphi _{d}/\partial t$:

\begin{equation}
\frac{\partial \mathbf{m}}{\partial \tau }=\mathbf{h\times m+}\lambda
\mathbf{m\times }\frac{\partial \mathbf{m}}{\partial \tau }.  \label{2}
\end{equation}%
Here%
\begin{equation}
\mathbf{h=}\frac{\partial ^{2}\mathbf{m}}{\partial \xi ^{2}}+(m_{z}+h_{0})%
\widehat{\mathbf{e}}_{z}+\varepsilon (\cos \varphi _{d}\widehat{\mathbf{e}}%
_{x}+\sin \varphi _{d}\widehat{\mathbf{e}}_{y}),  \label{3}
\end{equation}%
and we use normalized magnetization $\mathbf{m=M/}M$, dimensionless time $%
\tau =\beta Mgt$ and coordinate $\xi =\sqrt{\beta /\gamma }z$ ($g$, $\gamma $%
, and $\beta $ being the gyromagnetic ratio, the exchange constant, and the
anisotropy constant, respectively), while $h_{0}=H_{0}/(M\beta )$, $%
\varepsilon =a/(\beta Mg)$, $\varphi _{d}=-\int \Omega _{d}d\tau $, $\Omega
_{d}(\tau )=\omega _{d}/(\beta Mg)$, and $\lambda $ is the Gilbert damping
parameter. We seek spatially periodic solutions of Eq. (\ref{2}) and proceed
from the dissipationless version of this equation in polar coordinates ($%
m_{x}=\sin \theta \cos \varphi $, $m_{y}=\sin \theta \sin \varphi $, $%
m_{z}=\cos \theta $):%
\begin{equation}
\theta _{\tau }=\Phi _{\xi \xi }\sin \theta +2\Phi _{\xi }\theta _{\xi }\cos
\theta -\varepsilon \sin \Phi ,  \label{4}
\end{equation}%
\begin{equation}
\Phi _{\tau }=\left( -\frac{1}{\sin \theta }\theta _{\xi \xi }+\Phi _{\xi
}^{2}\cos \theta \right) +\cos \theta -\Omega _{d}^{\prime }(\tau
)-\varepsilon \cot \theta \cos \Phi .  \label{5}
\end{equation}%
where $\Phi =\varphi -\varphi _{d}$ is the phase mismatch and $\Omega
_{d}^{\prime }=\Omega _{d}-h_{0}$. This system is a spatial generalization
of the recently studied autoresonant magnetization switching problem in
single-domain nanoparticles \cite{Manfredi,Lazar145}, where one neglects the
spatial derivatives in Eqs. (\ref{4}) and (\ref{5}) to get
\begin{eqnarray}
\theta _{\tau } &=&-\varepsilon \sin \Phi ,  \label{6} \\
\Phi _{\tau } &=&\cos \theta -\Omega _{d}^{\prime }(\tau )-\varepsilon \cot
\theta \cos \Phi .  \label{7}
\end{eqnarray}%
In the 1D ferromagnetic case, Eqs. (\ref{6}), (\ref{7}) describe a spatially
uniform, rotating around the axis magnetization dynamics. In the rest of
this section, we discuss formation and stability of autoresonant uniform
states in the dissipationless case, but include dissipation in numerical
simulations for comparison.

The autoresonance idea is based on a self-sustained phase-locking of the
driven nonlinear system to chirped frequency driving perturbation. Typically
this phase-locking is achieved by passage through resonance with some
initial equilibrium. In our case, we assume linearly chirped driving
frequency $\Omega _{d}^{\prime }(\tau )=1-\alpha \tau $ for simplicity,
proceed from $\theta \approx 0$ $(m_{z}=1)$ at large negative time and
slowly pass the resonance $\Omega _{d}^{\prime }=1$ at $\tau =0$. For small $%
\theta $ Eqs. (\ref{6}), (\ref{7}) can be written as
\begin{eqnarray}
\frac{d\theta }{d\tau } &=&-\varepsilon \sin \Phi ,  \label{8} \\
\theta \frac{d\Phi }{d\tau } &=&\left( \alpha \tau -\theta ^{2}/2\right)
\theta -\varepsilon \cos \Phi ,  \label{9}
\end{eqnarray}%
which can be transformed into a single complex equation for $\Psi =\theta
e^{i\Phi }$%
\begin{equation}
i\frac{d\Psi }{d\tau }+(\alpha \tau -\left\vert \Psi \right\vert ^{2}/2)\Psi
=\varepsilon .  \label{10}
\end{equation}%
This NLS-type equation was studied in many applications and yields efficient
phase locking at $\Phi \approx \pi $ after passage through linear resonance
at $\tau =0,$ provided $\varepsilon $ exceeds a threshold \cite{Scholarpedia}
\begin{equation}
\varepsilon _{th}=0.58\alpha ^{3/4}.  \label{10b}
\end{equation}%
Later (for $\tau >0$), the phase locking continues as the nonlinear
frequency shift follows that of the driving frequency, i.e. $\theta
^{2}/2\approx \alpha \tau $. Importantly, this continuing phase-locking is
characteristic of any variation of the driving frequency [then $\alpha $ in (%
\ref{10}) represents the local frequency chirp rate at the initial
resonance], while the system remains in an approximate nonlinear resonance
\begin{equation}
m_{z}=\cos \theta \approx \Omega _{d}^{\prime }(\tau ),  \label{10a}
\end{equation}%
as long as the driving frequency chirp rate remains sufficiently small.
Under these conditions, the magnetization angles $\theta $ and $\varphi
\approx \varphi _{d}+\pi $ are efficiently controlled by simply varying the
driving frequency.
\begin{figure}[tp]
\centering\includegraphics[width=8.8cm]{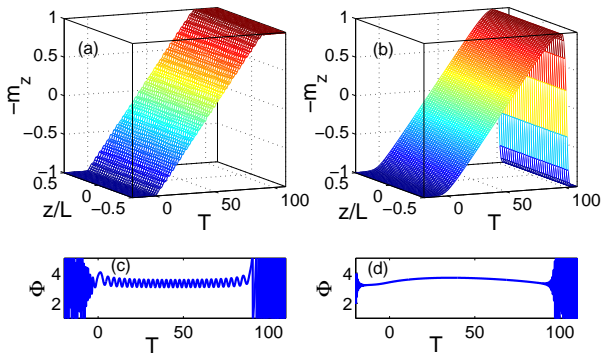}
\caption{(Color online) The uniform autoresonant magnetization state. (a)
-z-component of magnetization $-m_{z}$ versus slow time $T=\protect\alpha %
^{1/2}\protect\tau $; (c) phase mismatch $\Phi (0,T)=\protect\phi -\protect%
\phi _{d}$. In both panels $\protect\lambda =0$ and $\protect\varepsilon %
=3\times 10^{-3}$. Panels (b) and (d) are the same as (a) and (c), but $%
\protect\lambda =3\times 10^{-3}$ and $\protect\varepsilon =3\times 10^{-2}$.
}
\label{Fig1}
\end{figure}
We illustrate this effect in Fig. 1, showing the results of numerical
simulations of the original system (\ref{2}), assuming spatial periodicity
of length $l=6$ and linearly chirped frequency $\Omega _{d}^{\prime }(\tau
)=1-\alpha \tau $. The initial conditions $\theta =0.01\left\vert \cos
(\kappa \xi )\right\vert $ ($\kappa =2\pi /l$) represented a small spatial
perturbation for studying stability of the uniform state and we used
parameters $\lambda =0$, $h_{0}=5$, $\varepsilon =3\times 10^{-3}$, and $%
\alpha =5\times 10^{-4}$ in Figs. 1a and 1c, while $\lambda =3\times 10^{-3}$
and $\varepsilon =3\times 10^{-2}$ in Figs. 1b and d. Our numerical scheme used
an equivalent system of two coupled NLS-type equations based on the quantum
two-level analog \cite{Feynman,Lazar145} described in the Appendix. Figures
1a,b (without and with damping, respectively) show the evolution of $%
-m_{z}=-\cos \theta $ versus slow time $T=\alpha ^{1/2}\tau $, which
approximately follows the linear time dependence $\cos \theta \approx \Omega
_{d}^{\prime }$ on time, while Figs. 1c,d represents the corresponding phase
mismatch $\Phi (0,T)=\varphi (0,T)-\varphi _{d}(T)$, and illustrate the
continuing azimuthal phase-locking in the system at $\Phi \approx \pi $.
Note that the uniform solution in this case is stable with respect to
spatial perturbations. The dissipation changes the threshold condition for
entering the autoresonant uniform state \cite{Lazar94,Lazar145}, has some
effect on the phase mismatch (compare Figs. 1c and d) and leads to the
collapse of the solution to the initial equilibrium after dephasing.
Nevertheless, in the phase-locked stage the autoresonant uniform solutions
are similar with and without damping and remains stable with respect to
spatial perturbations. In contrast to the example in Fig. 1, one observes a
spatial instability of the autoresonant uniform state in Figs. 2a and b,
showing the numerical simulations with the same parameters as in Fig. 1a and
b, but $l=8$ instead of $6$. One can see the destruction of the uniform
state in Fig. 2 and formation of a complex spatio-temporal structure of $%
m_{z}(\xi /l,T)$ starting $T\approx 21$ in Fig. 1a and somewhat earlier in
Fig. 1b. These results can be explained by a perturbation theory as
described below.
\begin{figure}[bp]
\centering \includegraphics[width=8.8cm]{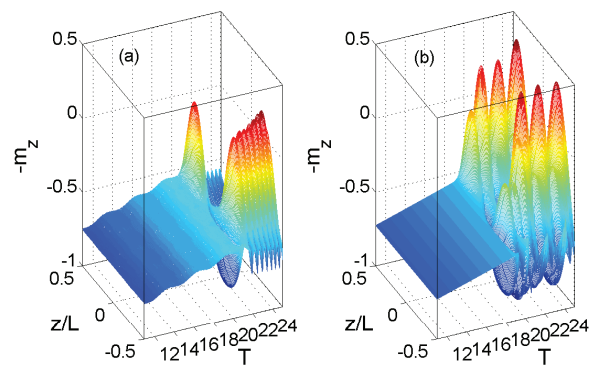}
\caption{(Color online) The instability of the uniform magnetization state.
The parameters of the simulations in (a) and (b) are the same as in Figs. 1a
and 1b respectively, but $\protect\kappa =2\protect\pi /L<1$. A complex
spatio-temporal magnetization profile develops beyond the point of
instability.}
\label{Fig2}
\end{figure}
We neglect damping for simplicity, freeze the time at $\tau =\tau _{0}$ and
set $\Phi =\pi +\delta \Phi $ and $\theta =\theta _{0}+\delta \theta $,
where $\theta _{0}$ satisfies
\begin{equation}
\cos \theta _{0}-\Omega _{d}^{\prime }(\tau _{0})+\varepsilon \cot \theta
_{0}=0.  \label{11}
\end{equation}%
Then, for small perturbations $\delta \Phi $ and $\delta \theta $ of
frequency $\nu $ and wave vector $\kappa $, Eqs. (\ref{4}) and (\ref{5})
become
\begin{eqnarray}
-i\nu \delta \theta  &=&-(\kappa ^{2}\sin \theta _{0}-\varepsilon )\delta
\Phi ,  \label{12} \\
-i\nu \delta \Phi  &=&\left( -\sin \theta _{0}+\frac{\kappa ^{2}}{\sin
\theta _{0}}-\frac{\varepsilon }{\sin ^{2}\theta _{0}}\right) \delta \theta ,
\notag
\end{eqnarray}%
yielding%
\begin{equation}
\nu ^{2}=\frac{1}{\sin ^{2}\theta _{0}}(\kappa ^{2}\sin \theta
_{0}-\varepsilon )\left( -\sin ^{3}\theta _{0}+\kappa ^{2}\sin \theta
_{0}-\varepsilon \right) .  \label{13}
\end{equation}%
One can see that for small $\varepsilon $, the uniform solution is stable
with respect to spatial perturbations provided%
\begin{equation}
\kappa >\sin \theta _{0}.  \label{14}
\end{equation}%
The examples in Figs. 1 ($\kappa =1.047$) and 2 ($\kappa =0.785$) are
consistent with this result.

\section{III Transformation from spatially uniform solution to a standing
wave}

\begin{figure}[tp]
\centering \includegraphics[width=8.8cm]{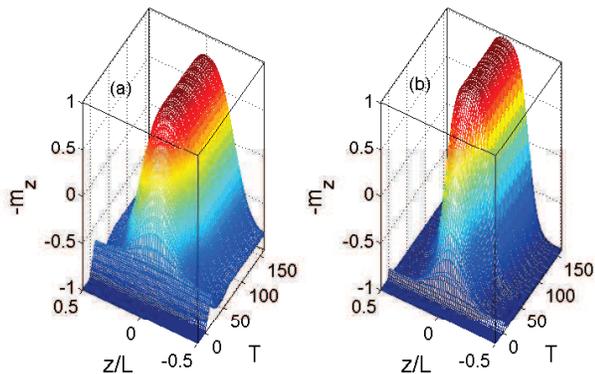}
\caption{(Color online) The formation of autoresonant standing waves from
the uniform magnetization state: (a) $\protect\kappa =0.78$, (b) $\protect%
\kappa =0.45$. The final waveform is reached as the driving frequency
gradualy decreases and stays constant for $T>69$.}
\label{Fig3}
\end{figure}
The formation of\ a uniform autoresonant solution $\cos \theta _{0}(\tau
)\approx $ $\Omega _{d}^{\prime }(\tau )$ in the spatially periodic LLG
problem was demonstrated above using a constant amplitude chirped frequency
drive, yielding stable evolution provided the inequality (\ref{14}) is
satisfied (see Fig. 1). If during the evolution, this inequality is
violated, the spatial instability develops (see Fig. 2). However, one can
avoid the instability and transform the uniform autoresonant solution into
an autoresonant standing wave by adding a simple spatial modulation of the
driving amplitude, i.e., uses $\varepsilon =\varepsilon _{0}+\varepsilon
_{1}\cos (\kappa \xi )$. We illustrate this phenomenon via simulations in
Fig. 3, where we use parameters $\alpha =5\times 10^{-4}$ and $\lambda =0$,
but, in the driving term, apply a modulated drive with $\varepsilon
_{1}=\varepsilon _{0}=3\times 10^{-3}$ and switch on $\varepsilon _{1}$ at $%
\tau =0$. The chirped driving frequency in this numerical example is of form
$\Omega _{d}^{\prime }=1-\Delta \Omega \sin (\alpha \tau /\Delta \Omega )$
for $\tau <\pi \Delta \Omega /2\alpha $ and $\Omega _{d}^{\prime }=1-\Delta
\Omega $ for $\tau >\pi \Delta \Omega /2\alpha $, and we use $\Delta \Omega
=0.98$. Thus, as in previous illustrations, the frequency passes the
resonance at $\tau =0$ having chirp rate $\alpha $, but then gradually
decreases reaching a constant. Figure 3a (where we use $l=8$) shows that the
addition of the spatial modulation of the driving amplitudes prevents the
spatial instability and leads to the emergence of a growing amplitude
standing wave solution. Figure 3b (where $l=13)$ shows a similar dynamics,
yielding formation of larger amplitude standing wave, which starts earlier,
at $T\approx 5$ (we again use the slow time $T=\alpha ^{1/2}\tau $ in this
and the following figures). The excited standing wave is fully controlled by
the variation of the driving frequency and precesses azimuthaly with the
angular velocity of the driving phase (due to the continuing phase locking
of $\Phi \approx \pi $). Furthermore, the magnetization waveform is
spatially locked to the driving perturbation, while the wave amplitude and
form is controlled by the instantaneous frequency of the drive. Importantly,
as $l$ increases, the maximum and the minimum of the final solution for $%
m_{z}$ become near $+1$ and $-1$, respectively. We have also verified
numerically that this solution approaches the well known soliton form with
exponentially falling tails [see Eq. (6.21) in Ref. \cite{Kosevich}]. We
further illustrate the autoresonant control of the standing magnetization
waves in Figs. 4a and 4c, where we show the results of simulations with all
the parameters of Fig. 3b, but instead of saturating the driving frequency,
allow it to vary according the same sinusoidal formula for an additional
time interval $\pi \Delta \Omega /2\alpha <\tau <\pi \Delta \Omega /2\alpha $%
, so the frequency returns to its original value. Figures 4b and 4d show the
results of similar simulations with the same parameters as in Figs. 4a and
4c, but $\varepsilon =10^{-2}$ and $\lambda =10^{-3}$. One observes the
return of the magnetization to its initial uniform state, being continuously
phase locked (see Fig. 4c and 4d) to the drive with or without dissipation.
\begin{figure}[bp]
\centering \includegraphics[width=8.8cm]{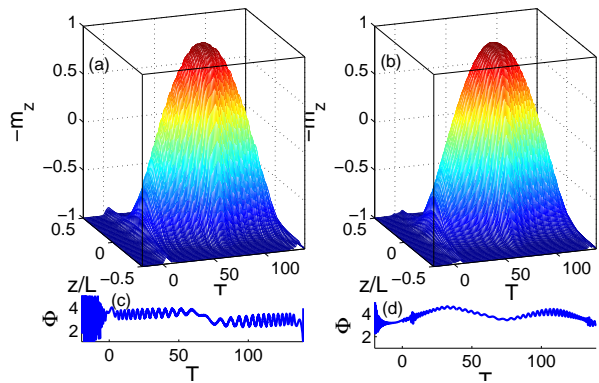}
\caption{(Color online) The control of the phase-locked standing
magnetization wave by varying the driving frequency. In (a) and (c) the
parameters of the simulations are the same as in Fig. 3b, but after reaching
its minimal value at $T=69$, the driving frequency increases back to the
initial value, while the magnetization returns to the initial state. Panels
(b) and (d) show $-m_{z}$ and phase mismatch $\Phi $, respectively versus
slow time in the same case as (a) and (c), but $\protect\lambda =10^{-3}$
and $\protect\varepsilon =10^{-2}$.}
\label{Fig4}
\end{figure}
The idea of the transformation from the uniform to standing wave solution by
passage through the spatial instability originates from the similarity to
the autoresonant excitations of standing waves of the driven-chirped
nonlinear Schrodinger (NLS)\ equation \cite{Lazar82}:%
\begin{equation}
i\psi _{\tau }+\psi _{\xi \xi }+\left\vert \psi \right\vert ^{2}\psi
+\varepsilon e^{-i\int \omega _{d}d\tau }=0.  \label{47}
\end{equation}%
If one writes $\psi =ae^{-i\phi }$ and separates the real and imaginary
parts in (\ref{47}), one arrives at the system%
\begin{eqnarray}
a_{\tau } &=&a\Phi _{\xi \xi }+2\Phi _{\xi }a_{\xi }-\varepsilon \sin \Phi ,
\label{48} \\
\Phi _{\tau } &=&-\frac{a_{\xi \xi }}{a}+\Phi _{\xi }^{2}-a^{2}-\omega
_{d}(t)-\frac{\varepsilon }{a}\cos \Phi .  \label{49}
\end{eqnarray}%
where $\Phi =\phi -\int \omega _{d}d\tau $. Similarly to our ferromagnetic
problem, the passage through the linear resonance in this system yields
excitation of the uniform autoresonant NLS solution followed by
transformation into autoresonant standing wave \cite{Lazar82}. One notices
the structural similarity between this NLS system and LLG Eqs.\ (\ref{4})
and (\ref{5}), so we proceed to the theory for the magnetization case using
the driven NLS ideas.

We assume that the time evolution in Eqs.\ (\ref{4}) and (\ref{5}) is slow
and interpret the solutions at a given time $\tau $, as being a slightly
perturbed solutions of the same system of equations, but with the time
derivatives and the forcing terms set to zero, i.e.,%
\begin{eqnarray}
\Phi _{\xi \xi }\sin \theta +2\Phi _{\xi }\theta _{\xi }\cos \theta &=&0,
\label{50} \\
\left( -\frac{1}{\sin \theta }\theta _{\xi \xi }+\Phi _{\xi }^{2}\cos \theta
\right) -\Omega _{d}^{\prime }(\tau )+\cos \theta &=&0.  \notag
\end{eqnarray}%
We notice that this is a dynamical, two degrees of freedom problem ($\xi $
serving as "time") governed by Hamiltonian
\begin{equation}
H=\frac{1}{2}(\theta _{\xi }^{2}+\Phi _{\xi }^{2}\sin ^{2}\theta )+V(\theta
),  \label{51}
\end{equation}%
where
\begin{equation}
V(\theta )=-\Omega _{d}^{\prime }(\tau )\cos \theta +\frac{1}{4}\cos
(2\theta ).  \label{52}
\end{equation}%
This fixed $\tau $ problem is integrable since it conserves the canonical
momentum $B=\Phi _{\xi }\sin ^{2}\theta $ and energy%
\begin{equation}
A=\frac{1}{2}\theta _{\xi }^{2}+V_{eff},  \label{53}
\end{equation}%
where $V_{eff}(\theta ,\tau )=\frac{B^{2}}{2\sin ^{2}\theta }+V(\theta )$.
Next, we discuss oscillating solutions of this problem and introduce the
conventional action-angle variables $(I,\Theta )$ and $(B,\Phi )$, where the
first pair describes pure $\theta $ oscillations in the effective potential $%
V_{eff}$, while the second pair is associated with the dynamics of $\Phi $.
If one returns to the original (time dependent and driven) system \ (\ref{4}%
) and (\ref{5}), $A(\tau )$ and $B(\tau )$ become slow functions of time. We
will present a theory describing these slow parameters via Whitham's average
variational principle \cite{Whitham} in the next section and devote the
remaining part of the current section to a simple qualitative picture of the
dynamics.
\begin{figure}[tp]
\centering \includegraphics[width=8.8cm]{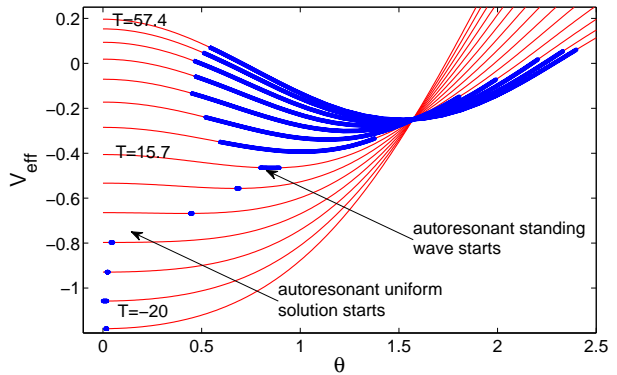}
\caption{(Color online) The formation of the autoresonant standing wave
modelled via dynamics of a quasi-particle in a slowly varying effective
potential $V_{eff}$. $V_{eff}$ versus $\protect\theta $ is shown for
successive times (thin red lines) starting at $T=-20$. The thick blue lines
show spatial oscillations of $\protect\theta $ at these times, as obtained
in simulations in Fig. 3a. The excitation proceeds as the quasi-particle
remains at the bottom of the potential well continuously, corresponding to
the flat solution. After passage through resonance with the spatial
modulation of the driving amplitude, autoresonant oscillations of the
quasi-particle in the effective potential are excited, describing the
standing magnetization wave.}
\label{Fig5}
\end{figure}
Our qualitative picture is based on the assumption of \textit{almost} purely
$\theta $ dynamics in the problem, i.e., setting $B\approx 0,$ which means a
continuous phase locking $\Phi \approx \pi $, simplifying the effective
potential to $V_{eff}\approx -\Omega _{d}^{\prime }(\tau )\cos \theta +\frac{%
1}{4}\cos (2\theta )$. As already discussed above, the phase locking at $\pi
$ is guaranteed in the initial excitation stage via temporal autoresonance
with constant amplitude $\varepsilon =\varepsilon _{0}$, chirped frequency
perturbation. But now our driving amplitude $\varepsilon =\varepsilon
_{0}+\varepsilon _{1}\cos (\kappa \xi )$ has two terms, where the first
leads to excitation of the uniform autoresonant solution as discussed above,
while the second term yields transition to the standing wave solution.
Initially, $\theta $ is efficiently trapped at the minimum location $\theta
_{m}$ of the potential well $V_{eff}$ given by $\cos \theta _{m}=\Omega
_{d}^{\prime }(\tau ).$ To $O(\varepsilon )$ this yields $\theta \approx
\theta _{m}$, so this dynamics corresponds to the uniform autoresonant
solution [see Eq. (\ref{11})]. The second term $\varepsilon _{1}\cos (\kappa
\xi )$ in the driving has little effect on the evolution at this stage,
until the spatial frequency $\kappa _{0}=\sqrt{\partial ^{2}V_{eff}/\partial
\theta _{m}^{2}}$ of oscillations of $\theta $ around $\theta _{m}$ passes
the resonance with this driving term, i.e. when%
\begin{equation}
\Omega _{d}^{\prime }(\tau )\cos \theta _{m}-\cos (2\theta _{m})\approx \sin
^{2}\theta _{m}=\kappa ^{2}.  \label{54}
\end{equation}%
But this is exactly the location of the instability of the uniform solution
[see Eq. (\ref{14})] without the term $\varepsilon _{1}\cos (\kappa \xi )$
in the drive. The passage through the resonance with this new drive term
excites growing amplitude oscillations of $\theta $ in the effective
potential. After the passage, the oscillations of $\theta $ become
autoresonant as the amplitude increases to preserve their spatial frequency
near $\kappa $ continuously. These newly induced spatially phase-locked,
growing amplitude oscillations of $\theta $ comprise the autoresonant
standing wave solution. The amplitude of these oscillations \ does not grow
indefinitely. Indeed, when the potential $V_{eff}$ becomes shallower again
as $\theta _{m}$ passes $\pi /2$ at $\Omega _{d}^{\prime }(\tau )=0$, the
spatial resonance can not be sustained, and the autoresonance is expected to
interrupt. We illustrate this dynamics in Fig. 5, showing the effective
potential $V_{eff}$ (thin red lines) at $14$ successive values of slow time
starting $T=-20$. The thick blue lines in the figure show the value of the
potential at $\theta (\xi ,T)$ at these times, as obtained in the
simulations in the example in Fig. 3a.

\section{IV Whitham's averaged variational analysis}

\subsection{Averaged Lagrangian density}

The LLG problem governed by Eqs.\ (\ref{4}) and (\ref{5}) allows Lagrangian
formulation with the Lagrangian density $L=L_{0}+L_{1}$ where
\begin{eqnarray}
L_{0}(\theta ,\theta _{\xi },\Phi _{\tau },\Phi _{\xi },\tau ) &=&\frac{1}{2}%
(\theta _{\xi }^{2}+\Phi _{\xi }^{2}\sin ^{2}\theta )+\Phi _{\tau }\cos
\theta  \notag \\
&&+\Omega _{d}^{\prime }(\tau )\cos \theta -\frac{1}{4}\cos (2\theta )
\label{21aa}
\end{eqnarray}%
and the perturbing part
\begin{equation}
L_{1}=-\varepsilon \sin \theta \cos \Phi .  \label{21a}
\end{equation}%
For studying the slow autoresonant evolution in system (\ref{4}) and (\ref{5}%
), we use Whitham's averaged Lagrangian approach. Following \cite%
{Lazar81,Lazar82}, describing a similar NLS problem, we seek solutions of
form
\begin{equation}
\theta =\vartheta (\tau )+U(\Theta ,\tau ),\Phi =\upsilon (\tau )+V(\Theta
,\tau ),  \label{22b}
\end{equation}%
where the explicit time dependence is slow, while $\Theta (\xi ,\tau )$ is a
fast variable and $U$ and $V$ are $2\pi $ periodic in $\Theta $. In
addition, the frequencies $\Theta _{\tau }=-\Omega (\tau )$ and $\beta
=\upsilon _{\tau }$ are slow functions of time and the wave vector $\Theta
_{\xi }=\kappa =const$ ($\kappa =2\pi /l$, $l$ being the periodicity length
in our problem). The Whitham's averaging \cite{Whitham} in this system
proceeds from the unperturbed Lagrangian density $L_{0}$, where one \textit{%
freezes }the slow time dependence at some\ $\tau $ and, using $\Phi _{\xi
}=\kappa V_{\Theta }$, replaces $\Phi _{\tau }=\beta -\Omega V_{\Theta
}=\beta -(\Omega /\kappa )\Phi _{\xi }$. This yields
\begin{eqnarray}
L_{0} &=&\frac{1}{2}(U_{\xi }^{2}+\Phi _{\xi }^{2}\sin ^{2}\theta )
\label{23} \\
&&+\left( \beta +\Omega _{d}^{\prime }-\frac{\Omega }{\kappa }\Phi _{\xi
}\right) \cos \theta -\frac{1}{4}\cos (2\theta ).
\end{eqnarray}%
Recall that the explicit dependence on $U$ in (\ref{23}) enters via $\theta
=\vartheta +U$. This Lagrangian density describes a two degrees of freedom
dynamical problem (for $U$ and $\Phi $), where $\xi $ plays the role of
"time". In dealing with this problem we use Hamiltonian formulation. We
define the usual canonical momenta
\begin{eqnarray}
P^{U} &=&\partial L_{0}/\partial U_{\xi }=U_{\xi }  \label{24} \\
P^{\Phi } &=&\partial L_{0}/\partial \Phi _{\xi }=\Phi _{\xi }\sin
^{2}\theta -\frac{\Omega }{\kappa }\cos \theta  \label{24a}
\end{eqnarray}%
and observe that $\Phi $ is a cyclic variable and therefore $P^{\Phi }=B$ is
the integral of motion. We will unfreeze the slow time dependence later and $%
B(\tau )$ will becomes a slow function of time. The Lagrangian density $%
L_{0} $ yields the Hamiltonian in the time-frozen problem
\begin{equation}
H_{0}=P^{U}U_{\xi }+P^{\Phi }\Phi _{\xi }-L_{0}  \label{24b}
\end{equation}%
and, after some algebra,%
\begin{equation}
H_{0}=H_{0}^{\prime }(P^{\theta },\theta )+V_{1}(\theta ,B,\Omega ,\beta ),
\label{25}
\end{equation}%
where%
\begin{eqnarray}
H_{0}^{\prime }(P^{U},U) &=&\frac{1}{2}\left( P^{U}\right) {}^{2}+V(\theta
)-V(\vartheta ),  \label{25a} \\
V(\theta ) &=&-\Omega _{d}^{\prime }\cos \theta +\frac{1}{4}\cos (2\theta ),
\label{25aa}
\end{eqnarray}%
and%
\begin{equation*}
V_{1}(B,\Omega ,\beta ,\theta )=V(\vartheta )+\frac{\left( B+\frac{\Omega }{%
\kappa }\cos \theta \right) ^{2}}{2\sin ^{2}\theta }-\beta \cos \theta .
\end{equation*}

At this stage, we return to the full driven (still time-frozen) problem
governed by the Hamiltonian%
\begin{equation}
H=H_{0}^{\prime }(P^{U},U)+V_{1}-L_{1}  \label{25d}
\end{equation}%
(recall that $L_{1}=-[\varepsilon _{0}+\varepsilon _{1}\cos (\kappa \xi
)]\sin \theta \cos \Phi $) and make canonical transformation from $P^{U},U$
to the action-angle ($AA$) variables $I,\Theta $ of Hamiltonian $%
H_{0}^{\prime }$. The dynamics governed by this Hamiltonian conserves its
energy $A=H_{0}^{\prime }$ and is periodic of period $2\pi $ in $\Theta $,
and, at this stage, we identify $\Theta $ with the angle variable used in
the definitions (\ref{22b}). The action variable in $H_{0}^{\prime }$
problem is%
\begin{equation}
I=\frac{1}{2\pi }\oint P^{U}dU=\frac{1}{2\pi }\oint \sqrt{2[A-V(\theta
)+V(\vartheta )]}dU,  \label{25e}
\end{equation}%
were the time dependence enters both explicitly in $V$ and via $\vartheta $.
Note that
\begin{equation}
\frac{\partial I}{\partial A}=\frac{1}{2\pi }\oint \frac{1}{\sqrt{%
2[A-V(\theta )+V(\vartheta )]}}dU=\frac{1}{\widetilde{\kappa }},  \label{25f}
\end{equation}%
$\widetilde{\kappa }(\vartheta ,A)$ being the (spatial) frequency of the
oscillations of $U$ governed by $H_{0}^{\prime }$. Next, we write the full
Lagrangian in our problem in terms of the new action angle variables%
\begin{equation}
L=\frac{d\Theta }{d\xi }I-H=\kappa I-A-V_{1}(B,\Omega ,\beta ,\theta
)+L_{1}(\theta ,\xi ,\tau ),  \label{25g}
\end{equation}%
where $\theta =\vartheta +U(I,\Theta ,\tau )$ in $V_{1}$ and $L_{1}$, as the
result of the canonical transformation. The Whitham's averaged Lagrangian
density $\Lambda $ is obtained by averaging $L$ in the time-frozen problem
over one oscillation governed by $H_{0}^{\prime }$:%
\begin{equation}
\Lambda =\frac{1}{2\pi }\int_{0}^{2\pi }Ld\Theta =\kappa I-A-\frac{1}{2\pi }%
\int_{0}^{2\pi }(V_{1}-L_{1})d\Theta .  \label{25h}
\end{equation}%
To complete the averaging, we calculate two remaining components\ $%
\left\langle V_{1}\right\rangle =\frac{1}{2\pi }\int_{0}^{2\pi }V_{1}d\Theta
$ and $\Lambda _{1}=\frac{1}{2\pi }\int_{0}^{2\pi }L_{1}d\Theta $ in (\ref%
{25h}).
\begin{eqnarray}
\left\langle V_{1}\right\rangle &=&V(\vartheta )+\frac{1}{2\pi }%
\int_{0}^{2\pi }\left[ \frac{\left( B+\frac{\Omega }{\kappa }\cos \theta
\right) ^{2}}{2\sin ^{2}\theta }-\beta \cos \theta \right] d\Theta  \notag \\
&=&V(\vartheta )+I_{1}\frac{B^{2}}{2}+I_{2}\frac{B\Omega }{\kappa }+I_{3}%
\frac{\Omega ^{2}}{2\kappa ^{2}}-\beta I_{4},  \label{26aa}
\end{eqnarray}%
where $I_{1}=\left\langle \frac{1}{\sin ^{2}\theta }\right\rangle
,I_{2}=\left\langle \frac{\cos \theta }{\sin ^{2}\theta }\right\rangle
,I_{3}=\left\langle \frac{\cos ^{2}\theta }{\sin ^{2}\theta }\right\rangle
,I_{4}=\left\langle \cos \theta \right\rangle $ and the averages $%
\left\langle ...\right\rangle $ are defined as%
\begin{equation}
\left\langle ...\right\rangle =\frac{1}{2\pi }\int_{0}^{2\pi }(...)d\Theta =%
\frac{\kappa }{2\pi }\oint \frac{(...)}{\sqrt{2[A-V(\theta )+V(\vartheta )]}}%
dU.  \label{26b}
\end{equation}%
Finally, we calculate the averaged driving part of the Lagrangian density
(recall that $\theta =\vartheta +U$ and $\Phi =\upsilon +V$)
\begin{equation}
\Lambda _{1}=-\frac{1}{2\pi }\int_{0}^{2\pi }\left[ \varepsilon
_{0}+\varepsilon _{1}\cos (\kappa \xi )\right] \sin (\vartheta +U)\cos
(\upsilon +V)d\Theta .  \label{26c}
\end{equation}%
Here, we limit evaluation of this averaged object to small spatial
oscillations of $\theta $ around $\vartheta $, write $U\approx a(I)\cos
\Theta $ and replace $\sin (\vartheta +U)\approx \sin \vartheta +a(I)\cos
\vartheta \cos \Theta $. Furthermore, we will also assume that $V$ is
sufficiently small to replace $\cos (\upsilon +V)\approx \cos \upsilon $.
Finally, assuming a continuous approximate double resonance in the problem,
i.e. $\upsilon (\tau )-\pi =\upsilon ^{\prime }\approx 0$ and $\Theta
-\kappa \xi -\pi =\mu (\tau )\approx 0$ (initial phase locking of $\upsilon $
at $\pi $ was shown in the uniform autoresonant solution stage), after
averaging

\begin{equation}
\Lambda _{1}\approx (\varepsilon _{0}\sin \vartheta -\frac{\varepsilon _{1}}{%
2}a(I)\cos \vartheta \cos \mu )\cos \upsilon ^{\prime }.  \label{27}
\end{equation}%
Therefore, our final averaged Lagrangian becomes%
\begin{eqnarray}
\Lambda &=&\kappa I-A-V(\vartheta )-\frac{I_{1}B^{2}}{2}-\frac{I_{2}B\Omega
}{\kappa }-\frac{I_{3}\Omega ^{2}}{2\kappa ^{2}}+I_{4}\beta  \notag \\
&&+\left[ \varepsilon _{0}\sin \vartheta -\frac{\varepsilon _{1}}{2}a(I)\cos
\vartheta \cos \mu \right] \cos \upsilon ^{\prime }.  \label{28a}
\end{eqnarray}%
We discuss the slow evolution of the full driven system next. Following
Whitham, this evolution is obtained by unfreezing the time and taking
variations of $\Lambda $ with respect to all dependent variables $%
A,\vartheta ,B,\Theta $ and $\upsilon $. Obviously, only slow objects enter
the averaged Lagrangian density.

\subsection{Evolution equations and stability analysis}

At this stage, we write variational evolution equations. The variation of $%
\Lambda $ with respect to $B$ yields%
\begin{equation}
\frac{d\mu }{d\tau }=-\Omega =\frac{\kappa I_{1}}{I_{2}}B,  \label{30}
\end{equation}%
and the variation with respect to $\Theta $ and use of (\ref{30}) results in

\begin{equation}
\frac{d}{d\tau }\left[ \left( I_{2}^{2}-I_{1}I_{3}\right) \frac{B}{\kappa
I_{2}}\right] \approx \frac{\varepsilon _{1}}{2}a\cos \vartheta \sin \mu .
\label{31a}
\end{equation}%
Similarly, the variation with respect to $A$ and $\upsilon $ gives%
\begin{eqnarray}
I_{4A}\frac{d\upsilon ^{\prime }}{d\tau } &=&1-\frac{\kappa }{\widetilde{%
\kappa }}+B^{2}\left( \frac{I_{1A}}{2}-\frac{I_{2A}I_{1}}{I_{2}}+\frac{%
I_{3A}I_{1}^{2}}{2I_{2}^{2}}\right)  \label{29} \\
&&+\frac{\varepsilon _{1}}{2}a_{A}\cos \vartheta \cos \mu \cos \upsilon
^{\prime }  \notag
\end{eqnarray}%
and
\begin{equation}
\frac{dI_{4}}{d\tau }\approx -(\varepsilon _{0}\sin \vartheta -\frac{%
\varepsilon _{1}}{2}a\cos \vartheta \cos \mu )\sin \upsilon ^{\prime }
\label{32}
\end{equation}%
Finally, the variation with respect to $\vartheta $ yields%
\begin{eqnarray}
I_{4\vartheta }\frac{d\upsilon ^{\prime }}{d\tau } &=&\frac{\partial
V(\vartheta )}{\partial \vartheta }-\kappa \frac{\partial I}{\partial
\vartheta }+B^{2}\left( \frac{I_{1\vartheta }}{2}-\frac{I_{2\vartheta }I_{1}%
}{I_{2}}+\frac{I_{3\vartheta }I_{1}^{2}}{2I_{2}^{2}}\right)  \notag \\
&&-(\varepsilon _{0}\cos \vartheta +\frac{\varepsilon _{1}}{2}a\sin
\vartheta \cos \mu )\cos \upsilon ^{\prime }.  \label{32cc}
\end{eqnarray}%
Equations (\ref{30})-(\ref{32cc}) comprise a complete set of slow evolution
equations for $A,B,\mu ,\upsilon ^{\prime }$ and $\vartheta $. The solution
of these equations proceeds by defining a quasi-steady state $B_{0}=\mu
_{0}= $ $\upsilon _{0}^{\prime }=0$, $\vartheta =\vartheta _{0}$ and $A_{0}$
given by
\begin{equation}
\left( V_{\vartheta }-\kappa I_{\vartheta }\right) _{A_{0},\vartheta
_{0}}-\varepsilon _{0}\cos \vartheta _{0}-\frac{\varepsilon _{1}}{2}a\sin
\vartheta _{0}=0,  \label{32e}
\end{equation}%
\begin{equation}
G(A_{0},\vartheta _{0})=\left( 1-\frac{\kappa }{\widetilde{\kappa }}+\frac{%
\varepsilon _{1}}{2}a_{A}\cos \vartheta _{0}\right) _{A_{0},\vartheta
_{0}}=0,  \label{32a}
\end{equation}%
Note that in the case $\varepsilon _{1}=0$ and small $A$, Eq. (\ref{32e})
nearly coincides with Eq. (\ref{11}) describing the autoresonant uniform
solution. Furthermore, for small $A,$ to $O(\varepsilon )$, Eq. (\ref{32e})
yields $V_{\vartheta _{0}}\approx 0$, i.e. $\vartheta _{0}$ remains near the
location of the minimum of $V(\theta )$ given by $\cos \vartheta _{0}\approx
\Omega _{d}^{\prime }$, as was suggested in the qualitative model in Sec. IV
and seen in simulations. On the other hand, Eq. (\ref{32a}) clarifies the
phase locking at $\mu \approx 0$ as $\widetilde{\kappa }$ approaches the
resonance $\widetilde{\kappa }=\kappa $ from below. For small perturbations $%
\delta A,\delta B,\delta \mu ,\delta \upsilon ^{\prime }$ and $\delta
\vartheta $ of the quasi-steady state, we use $I\approx \delta A/\widetilde{%
\kappa }\approx \delta A/\sin \vartheta $ to get the lowest order (linear)
set of equations%
\begin{eqnarray}
\frac{d\delta \mu }{d\tau } &=&\frac{\kappa I_{1}}{I_{2}}\delta B,
\label{36a} \\
\frac{d\delta B}{d\tau } &=&-\frac{\kappa \varepsilon _{1}aI_{2}\cos
\vartheta }{2(I_{1}I_{3}-I_{2}^{2})}\delta \mu ,  \label{34} \\
I_{4A}\frac{d\delta \upsilon ^{\prime }}{d\tau } &=&G_{\vartheta }\delta
\vartheta +G_{A}\delta A,  \label{35} \\
I_{4A}\frac{d\delta A}{d\tau }+I_{4\vartheta }\frac{d\delta \vartheta }{%
d\tau } &=&-(\varepsilon _{0}\sin \vartheta -\frac{\varepsilon _{1}}{2}a\cos
\vartheta )\delta \upsilon ^{\prime },  \label{36} \\
I_{4\vartheta }\frac{d\delta \upsilon ^{\prime }}{d\tau } &=&V_{\vartheta
\vartheta }\delta \vartheta -\kappa R\delta A  \label{36b}
\end{eqnarray}%
where we use $\widetilde{\kappa }\approx \sin \vartheta $, so $R\approx \cos
\vartheta /\sin ^{2}\vartheta $ and all coefficients in (\ref{36a}) - (\ref%
{36b}) are viewed as constants evaluated at the quasi-steady state. Eqs. (%
\ref{36a}) and (\ref{34}) yield%
\begin{equation}
\frac{d^{2}\delta \mu }{d\tau ^{2}}+\nu _{1}^{2}\delta \mu \approx 0,
\label{36c}
\end{equation}%
while Eqs. (\ref{35})-(\ref{36b}) reduce to
\begin{equation}
\frac{d^{2}\delta \upsilon ^{\prime }}{d\tau ^{2}}+\nu _{2}^{2}\delta
\upsilon ^{\prime }\approx 0,  \label{36d}
\end{equation}%
where the two frequencies satisfy
\begin{eqnarray*}
\nu _{1}^{2} &=&\frac{\varepsilon _{1}\kappa ^{2}I_{1}a\cos \vartheta }{%
2(I_{1}I_{3}-I_{2}^{2})}, \\
\nu _{2}^{2} &=&\frac{(\varepsilon _{0}\sin \vartheta -\frac{\varepsilon _{1}%
}{2}a\cos \vartheta )(G_{A}V_{\vartheta \vartheta }-\kappa RG_{\vartheta })}{%
I_{4A}(I_{4A}V_{\vartheta \vartheta }-\kappa RG_{\vartheta })+I_{4\vartheta
}(G_{A}I_{4\vartheta }-G_{\vartheta }I_{4A})}.
\end{eqnarray*}%
A positiveness of $\nu _{1,2}^{2}$ guarantees stability of the (doubly)
autoresonant ($\upsilon ^{\prime }\approx 0$ and $\mu \approx 0$) evolution
of the system. We observe that
\begin{equation*}
I_{1}I_{3}-I_{2}^{2}\propto \left( \sum_{i}\frac{1}{s_{i}}\right) \left(
\sum_{j}\frac{x_{j}^{2}}{s_{j}}\right) -\left( \sum_{i}\frac{x_{i}}{s_{i}}%
\right) ^{2},
\end{equation*}%
where $x_{i}=\cos \theta _{i}$ and $s_{i}=\sin ^{2}\theta _{i}\sqrt{%
[A-V_{eff}(\theta _{i})]}$ . Then%
\begin{equation}
I_{1}I_{3}-I_{2}^{2}\propto \sum_{i,j>i}\frac{(x_{i}-x_{j})^{2}}{s_{i}s_{j}},
\label{39}
\end{equation}%
so $\nu _{1}^{2}$ is positive for $\vartheta _{0}<\pi /2$. Then, since $%
B=\delta B$, and $\mu =\delta \mu $, they both remain small. Furthermore,
for small excitations of $A$, to lowest order in $A,$ $\kappa =\sin
\vartheta $, $G_{\vartheta }=\cos \vartheta /\sin \vartheta $, $G_{A}=(\frac{%
1}{\sin \vartheta }-\frac{3}{2\sin ^{3}\vartheta })$, $I_{4\vartheta }=-\sin
\vartheta $, $I_{4A}=\cos \vartheta /\sin ^{2}\vartheta $. With these
substitutions, one finds $\nu _{2}^{2}\approx \varepsilon _{0}\sin \vartheta
-\frac{\varepsilon _{1}}{2}a\cos \vartheta $. Then condition $\varepsilon
_{0}\sin \vartheta -\frac{\varepsilon _{1}}{2}a\cos \vartheta >0$ guarantees
the stability of the autoresonant evolution.

\section{V Spin torque driving}

\begin{figure}[bp]
\centering \includegraphics[width=8.8cm]{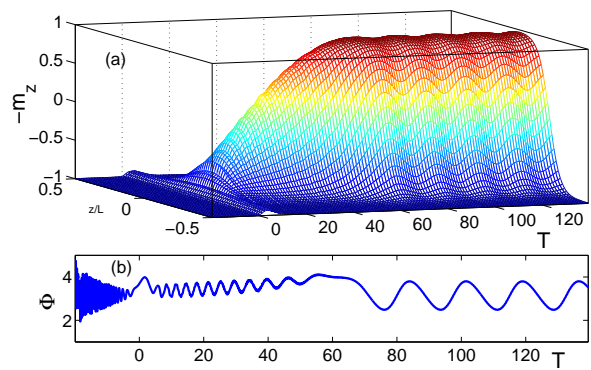}
\caption{(Color online) Formation of the autoresonant standing magnetization
wave by chirped frequency spin torque drive. The parameters in the
simulations are the same as in Fig. 3b for the AC drive. Panels (a) and (b)
show $-m_{z}$ and phase mismatch $\Phi(0,T)$, respectively versus $T$. }
\label{Fig6}
\end{figure}
The excitation of the uniform solution and its transition to the\
autoresonant standing wave can be also achieved by using spin torque drive
instead of the microwave drive used above (a related autoresonant problem
for single domain nano-particles was studied in Ref. \cite{Lazar152}). The
effective magnetic field associated with the spin torque is%
\begin{equation}
\mathbf{h}_{s}=\mathbf{m\times I}_{s},  \label{60}
\end{equation}%
where $\mathbf{I}_{s}$ is the dimensionless spin polarized current, which
will be assumed of form $\mathbf{I}_{s}=2\varepsilon \sin \varphi _{d}%
\mathbf{e}_{x}$ in the following, yielding
\begin{equation}
\mathbf{h}_{s}=2\varepsilon (m_{z}\mathbf{e}_{y}-m_{y}\mathbf{e}_{z}).
\label{61}
\end{equation}%
The analog of system (\ref{4}), (\ref{5}) for this drive is

\begin{eqnarray}
\theta _{\tau } &=&\Phi _{\xi \xi }\sin \theta +2\Phi _{\xi }\theta _{\xi
}\cos \theta -\varepsilon \cos \theta \sin \Phi ,  \label{63} \\
\Phi _{\tau } &=&\left( -\frac{1}{\sin \theta }\theta _{\xi \xi }+\Phi _{\xi
}^{2}\cos \theta \right) -\Omega _{D}^{\prime }-\frac{\varepsilon \cos \Phi
}{\sin \theta },  \label{63a}
\end{eqnarray}%
where $\Phi =\varphi -\varphi _{d}+\pi /2$.

Note that for small $\theta $ the last two equations are nearly the same as
Eqs. (\ref{4}), (\ref{5}) for the microwave drive. One consequence of this
is that the autoresonance threshold when passing the linear resonance is the
same for both cases. Figure 6 illustrates the formation and control of the
autoresonant standing wave via a spin torque drive in simulations using the
parameters of Fig. 3b. One can see that the form of the excited solution in
Figs. 3b and 6a are very similar. Despite this similarity, a complete
Whitham's-type theory of the spin torque driven problem is more complex than
that for the microwave drive case, because the driving parts in Eqs. (\ref%
{63}) and (\ref{63a}) do not allow Lagrangian description. Therefore, we
leave this theory outside the scope of the present work.

\section{VI Conclusions}

In conclusion, we have studied the problem of autoresonant excitation and
control of 1D standing magnetization waves in easy axis ferromagnetics in an
external magnetic field \ and driven by a weak circularly polarized, chirped
frequency microwave field. We had modeled this problem by the spatially
periodic time dependent LLG equation [see Eq. (\ref{2})]. We had discussed
the excitation of the autoresonant solutions in this system via theory and
compared the results with numerical simulations. The excitation proceeded as
the driving frequency passed a resonance with the initially spatially
uniform magnetization equilibrium in the direction of the easy axis (polar
angle $\theta =0$), yielding a driven spatially uniform magnetization with
the azimuthal angle $\varphi $ of the magnetization locked (and therefore
controlled) by the phase of the microwave. This phase locking
(autoresonance) reflects a continuous self-adjustment of $\theta $ [ see Eq.
(\ref{10a})], so that the resonance is preserved despite the variation of
the driving frequency. It was shown that the condition for this autoresonant
evolution is the driving amplitude $\varepsilon $ exceeding a threshold,
which scales with the driving frequency chirp rate as $\varepsilon _{th}\sim
\alpha ^{3/4}$ [see Eq. (\ref{10b})]. We had also shown that the uniform
autoresonant magnetization state remained stable with respect to spatial
perturbations if $\sin \theta <\kappa =2\pi /l$, $l$ being the periodicity
length in the problem. In the case $2\pi /l>1$, the stable uniform state
reached a complete magnetization inversion ($\theta \rightarrow \pi $\.{)}.
In contrast, when $\theta $ increased during the autoresonant uniform state
evolution and passed the point where $\sin \theta =2\pi /l$, the spatial
instability developed, yielding a complex spatio-temporal magnetization wave
form.

We had shown that if instead of a constant driving amplitude, one introduced
a spatially modulated amplitude $\varepsilon _{0}+\varepsilon _{1}\cos
(\kappa \xi )$, then, instead of the instability, a growing amplitude
standing wave with the amplitude and form controlled by the frequency of the
driving wave. This emerging autoresonant solution is doubly phase locked,
i.e. its azimuthal angle $\varphi $ is locked to the phase of the driving
wave, while $\theta $ performs slowly evolving growing amplitude nonlinear
spatial oscillations in an effective potential, which are continuously
phase-locked to the spatial frequency $\kappa $ of the modulation of the
drive. Furthermore, as the periodicity length $l$ increases, the
autoresonant standing wave approaches the well know soliton form [see Eq.
(6.21) in \cite{Kosevich}]. The formation of the autoresonant standing is
fully reversible and can be returned to its initial uniform ($\theta \approx
0$) state by simply reversing the variation of the driving frequency. In
addition to suggesting a qualitative description of this autoresonant
evolution (see Sec. III), we had developed a complete theory of the dynamics
in the problem based on the Whitham's averaged variational approach and
studied modulational stability of the autoresonant solutions (see sec. IV).
We had found numerically that a sufficiently weak dissipation does not
affect the autoresonant evolution significantly. We had also discussed
formation of autoresonant standing waves when replacing the microwave drive
by a spatially modulated transverse spin torque driving and illustrated this
possibility in numerical simulations. Developing a full Whitham's type
theory in this case and inclusion of dissipation and thermal fluctuations in
the theory seem to be important goals for future research. Finally, it is
known that the undriven, dissipationless LLG problem (\ref{2}) is integrable
\cite{Kosevich}. This means that there exist many additional, so called
multiphase solutions in this problem. Addressing the question of excitation
and control of this multitude of solutions by chirped frequency
perturbations seems to comprise another interesting goal for the future.

\begin{acknowledgements}
The authors would like to thank J.M. Robbins and E.B. Sonin for stimulating discussions and important comments. This work was supported by the Israel Science Foundation
Grant No. 30/14 and the Russian state program AAAA-A18-118020190095-4.
\end{acknowledgements}

\section{Appendix: Quantum two-level model}

We perform our numerical simulations to lowest significant order in $\lambda
$ and, therefore, approximate LLG Eq. (\ref{2}) as%
\begin{equation}
\frac{\partial \mathbf{m}}{\partial \tau }\approx \mathbf{h\times m+}\lambda
\mathbf{m\times (h\times m)=h}^{\prime }\mathbf{\times m,}  \label{71}
\end{equation}%
where $\mathbf{h}^{\prime }=\mathbf{h}-\lambda \mathbf{h\times m.}$ Our
numerical scheme for studying the evolution governed by Eq. (\ref{71}) is
based on the equivalent quantum two-level system (idea originated by Feynman
\cite{Feynman}, and recently used in studying magnetization inversion in
single domain nano-particles \cite{Lazar145, Lazar152}). We solve
\begin{equation}
i\frac{\partial A_{1}}{\partial \tau }=\frac{d_{0}}{2}A_{1}+dA_{2},
\label{76}
\end{equation}

\begin{equation}
i\frac{dA_{2}}{\partial \tau }=-\frac{d_{0}}{2}A_{2}+d^{\ast }A_{1},
\label{77}
\end{equation}%
where $A_{1,2}=A_{1,2}(\xi ,\tau )$ are the wave functions of a pair of
coupled quantum levels and
\begin{eqnarray}
d_{0} &=&h_{z}^{\prime },  \label{76a} \\
d &=&\frac{(h_{x}^{\prime }-ih_{y}^{\prime })}{2}.  \label{76b}
\end{eqnarray}
The magnetization $\mathbf{m}$ in $d_{0\text{ }}$and $d$ in Eqs. (\ref{76}),
(\ref{77}) is related to $A_{1,2}$ via%
\begin{eqnarray}
m_{x} &=&A_{1}A_{2}^{\ast }+A_{1}^{\ast }A_{2}=2B_{1}B_{2}\cos \varphi ,
\notag \\
m_{y} &=&i(A_{1}A_{2}^{\ast }-A_{1}^{\ast }A_{2})=2B_{1}B_{2}\sin \varphi ,
\label{77a} \\
m_{z} &=&\left\vert A_{1}\right\vert ^{2}-\left\vert A_{2}\right\vert
^{2}=B_{1}^{2}-B_{2}^{2},  \notag
\end{eqnarray}%
where $A_{1,2}=B_{1,2}\exp (i\varphi _{1,2})$ and $\varphi =\varphi
_{2}-\varphi _{1}$. Note that, as expected, the total population of our two
level system remains constant, $\left\vert A_{1}\right\vert ^{2}+\left\vert
A_{2}\right\vert ^{2}=|\mathbf{m|}=1$. Note also that $m_{\bot }=\sqrt{%
m_{x}^{2}+m_{y}^{2}}=2B_{1}B_{2}$, while $\varphi $ is the azimuthal
rotation angle of the magnetization around $\xi $. Formally, the system (\ref%
{76}), (\ref{77}) comprises a set of two coupled NLS-type equations for wave
functions $A_{1,2}$. The numerical approach to solving this system
throughout this work used a standard pseudospectral method \cite{Canuto}
subject to given initial and periodic boundary conditions.

\end{document}